\documentclass{pasj00}
\draft

\begin{document}
\SetRunningHead{J. Takahashi et al.}{Earthshine Polarization Spectra}
\Received{2012/09/10}%{yyyy/mm/dd}
\Accepted{2012/11/10}%{yyyy/mm/dd}
%\Published{}%{yyyy/mm/dd}

\title{Phase Variation of Earthshine Polarization Spectra}

%%% begin:list of authors
% Do NOT capitalize all letters in "textsc".

%\author{Jun \textsc{Takahashi} %
%  \thanks{Example: Present Address is xxxxxxxxxx}}
%\affil{A-Address of Institute}
%\email{aaaaa@xxx.xxx.xx.xx}

%\author{B-Firstname \textsc{B-Familyname}}
%\affil{B-Address of Institute}\email{bbbbb@xxx.xxx.xx.xx}
%\and
%\author{C-Firstname {\sc C-Familyname}}
%\affil{C-Address of Institute}\email{ccccc@xxx.xxx.xx.xx}
%%% end:list of authors

%%% Please use the following style in case that sorting by 
%%% affiliation is impossible. 
%
\author{%
   Jun \textsc{Takahashi},\altaffilmark{1}
   Yoichi \textsc{Itoh},\altaffilmark{1}
   Hiroshi \textsc{Akitaya},\altaffilmark{2}
   Akira \textsc{Okazaki},\altaffilmark{3}
   Koji \textsc{Kawabata},\altaffilmark{2}
   Yumiko \textsc{Oasa},\altaffilmark{4} 
   and
   Mizuki \textsc{Isogai}\altaffilmark{5}}
\altaffiltext{1}{Nishi-Harima Astronomical Observatory, Center for Astronomy, University of Hyogo, 407-2 Nishigaichi, Sayo, Hyogo, 679-5313}
\email{takahashi@nhao.jp}
\altaffiltext{2}{Astrophysical Science Center, Hiroshima University, 1-3-1 Kagamiyama, Higashi-Hiroshima, Hiroshima 739-8526}
\altaffiltext{3}{Department of Science Education, Gunma University, 4-2 Aramaki, Maebashi, Gunma 371-8510}
\altaffiltext{4}{Faculty of Education, Saitama University, 255 Shimo-Okubo, Sakura-ku, Saitama 338-8570}
\altaffiltext{5}{Koyama Astronomical Observatory, Kyoto Sangyo University, Motoyama, Kamigamo, Kita-ku, Kyoto, Kyoto 603-8555}

%% `\KeyWords{}' always has to be placed before `\maketitle'.
 %Do NOT move this preamble from here!

\maketitle

\begin{abstract}
We present the results of the optical spectropolarimetry of Earthshine on the Moon for Earth phase angles ranging from 49$^{\circ}$ to 96$^{\circ}$. 
The observations were conducted on 2011 March 9-13 (UT) using the  spectropolarimeter HBS installed on the 1.88 m telescope at Okayama Astrophysical Observatory.
The wavelength coverage was 450-850 nm with a resolution of 6 nm.
The observed Earthshine polarization degree spectra exhibit decreasing polarization degree with increasing wavelength at any phase.
The overall degree of polarization increases as the Earth approaches a quadrature phase.
The phase dependence differs with the wavelengths; 
the maximum polarization for the $V$ band occurs at a phase angle of $\sim 90^\circ$, whereas that for longer wavelengths is reached at larger phase angles.
This is interpreted as indicating that Earthshine polarization at shorter wavelengths is dominated by atmospheric Rayleigh scattering, whereas that at  longer wavelengths has an increasingly effective contribution from the Earth surface reflection.
The wavelength dependence of the phase angle of the maximum polarization appears to be unique among the terrestrial planetary bodies in the Solar System.
Therefore this might constitute important evidence pointing toward a distinctive characteristic  of the Earth:  the planet has a scattering  but transparent atmosphere  above its surface.
\end{abstract}

\section{Introduction}\label{sec:intro}
The identification of an Earth-like planet will have a great impact on astronomy and science in general.
A growing number of discovered extrasolar planets are fostering anticipation for reaching the goal.
Indeed, several extrasolar planets, such as GJ 667C c \citep{angl2012} and Kepler-22 b \citep{boru2012}, are suspected of being Earth-like on the basis of their orbit and mass/size.
Such discoveries have spurred a demand for a methodology for  characterizing a candidate as Earth-like.
Because the polarization spectrum of planetary reflected light contains physical and compositional information about the planet, the spectropolarimetry of extrasolar planets has potential to be a strong tool for  characterizing a planet.
Ground-based polarimetry is advantageous because the degree of polarization is, in principle, not affected by telluric absorption.

\citet{stam2008} calculated the optical polarization spectra of light reflected by a modeled Earth-like exoplanet.
The model showed that the light reflected by an Earth-like exoplanet can be highly polarized by light scattering  in the  atmosphere, and  by surface reflection.
Single scattering generally produces higher polarization than multiple scattering or reflection from a rough surface. 
The degree of polarization in the continuum of the spectra studied by \citet{stam2008} decreased with increasing wavelength.
This was explained by an increasing contribution of surface reflection to the total planetary reflected light due to a decrease in the atmospheric absorption optical thickness. 
In addition, \citet{stam2008} derived the presence of enhanced polarization features at gaseous molecule absorption bands, e.g., the O$_2$-A band (760 nm).
The higher polarization was the result of a lower fraction of multiply scattered light (and thus a higher fraction of singly scattered light) in an absorption band as compared with the nearby continuum region.
\citet{stam2008} also examined the effect of cloud coverage and found that it reduced the polarization degree; this was attributed to an increased fraction of multiple scattering.

Different scattering or reflecting media exhibit different phase-angle dependences of the polarization. 
Hence, the phase angle at which the maximum polarization degree occurs ($\theta_{\rm{max}}$) can be used to characterize exoplanets \citep{zugg2010}.
For instance, $\theta_{\rm{max}}$ is 90$^{\circ}$ for molecular Rayleigh single scattering \citep{hapk1993}.
In contrast, $\theta_{\rm{max}}$ for a particulate surface can be larger than 90$^{\circ}$ \citep{hapk1993, stam2008, zugg2010}.
In fact, it is known that the lunar (airless solid) surface has a $\theta_{\rm{max}}$ of 95-110$^\circ$ \citep{coyn1970, doll1971, hapk1993}.

The Earth, the only ``known Earth-like planet'', should be observed carefully in advance to facilitate the identification of an Earth-like exoplanet.
Nonetheless there have been very few observational attempts to obtain the polarization spectrum of the Earth. 
The satellite-borne instrument POLDER conducted polarimetry of the Earth in three optical bands \citep{desc1994}.
However, its limited field of view (swath: 2400 km) should be accompanied by a challenge in reproducing the polarization of the disk-integrated Earthlight, which can be used as a direct comparison with a distant (and thus point-source) exoplanet. 

We conducted spectropolarimetry of Earthshine on the Moon as an alternative approach.
Earthshine is Earthlight reflected from the lunar surface.
It is commonly used by ground-based observers to obtain the disk-integrated light of the Earth because the spatial information in Earthlight is averaged out in the lunar surface reflection.
Although photometry (e.g., \cite{danj1928,dubo1947,qiu2003,pall2003} \& 2004; \cite{mont2007,lang2009}), regular intensity spectrometry (e.g., \cite{arno2002,wool2002,seag2005,mont2005} \& 2006; \cite{turn2006,hamd2006, pall2009}), and monochromatic polarimetry \citep{doll1957}  of Earthshine have been conducted (as reviewed by \cite{arno2008,pall2010}), spectropolarimetry had never been reported until the recent publication of \citet{ster2012}, who observed Earthshine when the Moon was at the waxing and waning quadratures.
Our observations were performed independently, and we successfully monitored  phase variations in the Earthshine polarization spectra.
The results allow us to discuss both the wavelength and phase dependence of the polarization, namely the wavelength dependence of $\theta_{\rm{max}}$; hence, they could provide more information than an observation at a single phase.

Note that in this paper, the light of the Earth as viewed from the space is referred to as \textit{Earthlight}, which is distinguished from \textit{Earthshine} on the Moon as observed from the ground.

\section{Observations}
\begin{figure}
  \begin{center}
    \FigureFile(160mm,120mm){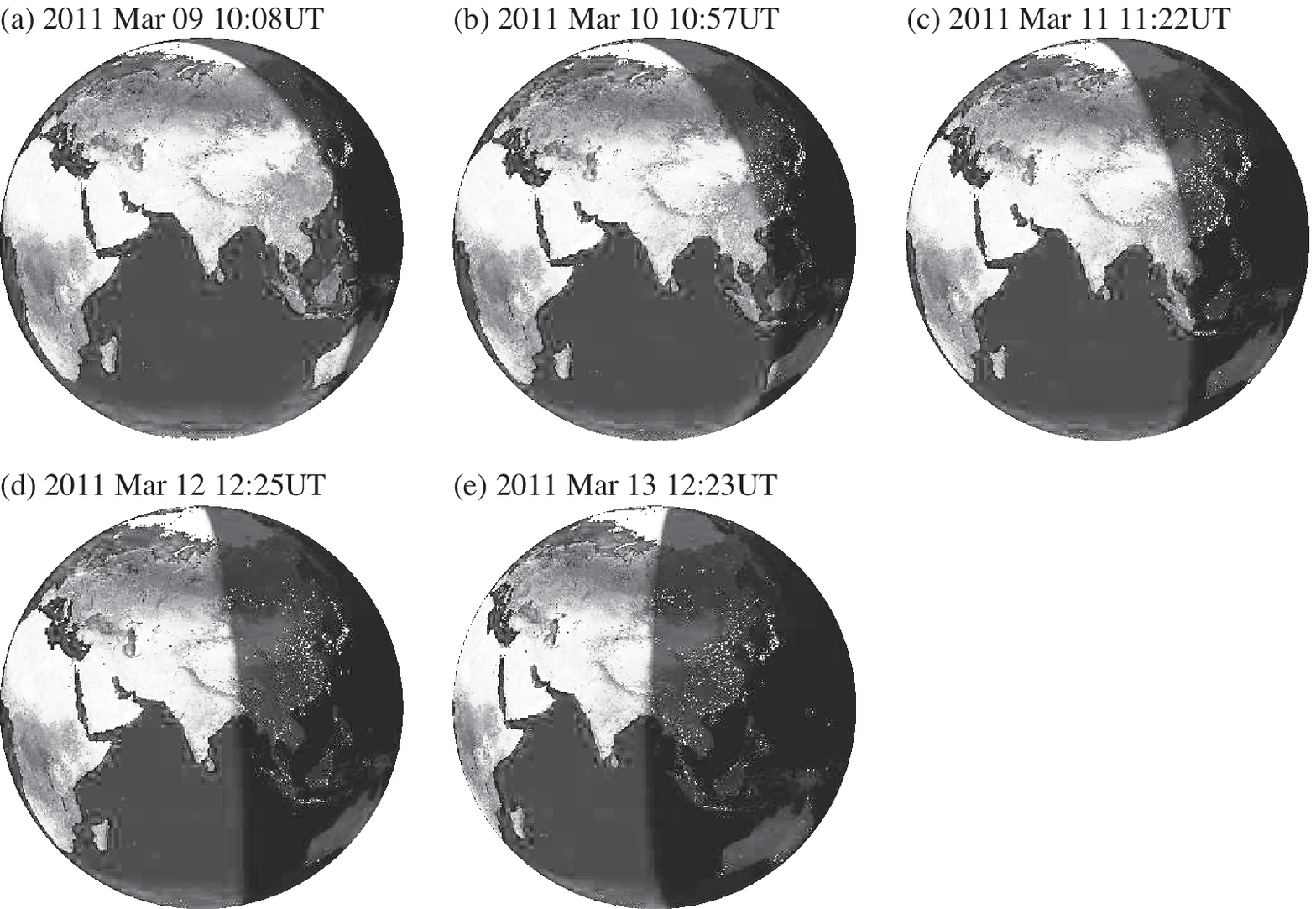}
  \end{center}
  \caption{Cloud-free Earth views as seen from the Moon at the midpoint of each observation. These images were produced with the Earth and Moon Viewer\footnotemark.}
  \label{fig:earthviews}
\end{figure}

Our observations were made on five different nights: 2011 March 9-13 (UT).
The Moon waxed from a new crescent to the first quarter.
The parameters of our Earthshine observations are summarized in Table \ref{tab:log}.
The Earth phase angle $\theta$, which is defined as the Sun-Earth-Moon angle in this paper, ranged from 49$^{\circ}$ to 96$^{\circ}$. 
The views of the Earth as seen from the Moon are reproduced in Fig.\ \ref{fig:earthviews}.
The contributions of different Earth scenes to Earthshine are shown in Table \ref{tab:surface}. 
The cloud coverage was virtually constant throughout the observation period.
The ratio of the contribution from the ocean to that from the land decreased from 1.05 to 0.51.

We used the spectropolarimeter HBS (\cite{kawa1999}) mounted at the Cassegrain focus (f/18) of the 1.88 m telescope at Okayama Astrophysical Observatory, located in Okayama, Japan.
The selected diaphragm was D2, which has twin rectangular holes aligned in the east-west direction.
Their size is 0.2 mm $\times$ 1.4 mm each, which corresponds to 1.2$''$ $\times$ 8.5$''$.
HBS employs a half-wave plate and a Wollaston prism.
A polarization spectrum is obtained by rotating the half-wave plate by 0$^{\circ}$, 22.5$^{\circ}$, 45$^{\circ}$, and 67.5$^{\circ}$.
In our observations, HBS provided polarization spectra at wavelengths of 450-850 nm with a resolution of 6 nm.
The attached CCD was cooled to $\sim -75 ^\circ$C.

An observational sequence consisted of Moonshine (MS), Earthshine (ES), near sky (NSKY), and far sky (FSKY) observations.
We obtained 1-4 sets of effective sequences each night.
For the MS spectra, the telescope was pointed to 12.9$'$ west of the Moon's center.
The use of MS is described in section \ \ref{sec:reduce}.
Then ES spectra were observed at 11.3$'$ east of the center.
NSKY and FSKY spectra for background subtraction were obtained by pointing toward the vicinity of the ES, which was located at 20$'$ and 30$'$ east of the Moon's center, respectively.
The single exposure time was 60-300 seconds for ES, 30 seconds for NSKY/FSKY, and 10-30 seconds for MS.

Dark frames were obtained every night.
The dark count was evaluated to be less than 1 ADU on the basis of a previous CCD performance test.
Two types of daylight flat frames were acquired to correct for the effect of the difference in the sizes of the twin holes on the diaphragm and of non-uniformity along the spatial axis of the spectral images\footnote{A detailed description is available in the HBS Reduction Manual (in Japanese).}.
The CCD pixel sensitivity correction was processed using a prepared master flat frame.
Stellar observations were made through a Glan-Taylor prism for instrumental depolarization correction.
Unpolarized and strongly polarized standard stars were observed for instrumental polarization correction.
Spectra from an Hg-Ne lamp were obtained once in the observation period for wavelength calibration.
The frame-by-frame wavelength shift was corrected using the absorption feature in the telluric O$_2$ A-band in the object frames. 

\begin{table}
  \caption{Parameters of Earthshine observations.
  Integration time is expressed as $exposure \times rotation \times set$.
  $\theta$ is the Earth phase angle (Sun-Earth-Moon angle).}
  \label{tab:log}
  \begin{center}
    \begin{tabular}{ccclc}
      \hline
        Date, 2011 & Start UT & End UT &
         Integration (s)& $\theta$ ($^\circ$)\\ 
      \hline
    Mar 09 &  9:55 & 10:21 &
     $300\times4\times1$ & 49 \\
	Mar 10 & 10:09 & 11:45&
	 $300\times4\times2 + 150\times4\times1$  & 60 \\
	Mar 11 & 10:18 & 12:25&	
	 $300\times4\times4$ & 72 \\
	Mar 12 & 11:03 & 13:48&
	 $300\times4\times4 + 60\times4\times1$  & 84 \\
	Mar 13 & 10:27 & 14:18 &
	 $300\times4\times4 + 150\times4\times1$ & 96 \\
      \hline
    \end{tabular}
  \end{center}
\end{table}

\begin{table}
  \caption{Contributions of different Earth scenes to Earthshine.
  The values are the mean within the time spans in the second column.
  TSDI in the fifth column stands for tundra$+$shrub$+$desert$+$ice.
  The values in a bracket in the sixth column are cloud fractions over the ocean and the land, respectively, where land means vegetation+TSDI.
  The final column is the ratio of the cloud-free ocean to the cloud-free land.
  These numbers have been calculated kindly by Dr. Enric Pall{\'e} using data taken from the MODIS experiment on board the Terra and Aqua satellites.
  These data have been put into a grid format and weighted by both the Sun-Earth and the Earth-Moon geometries.
  Further information can be found in \citet{pall2003} and \citet{mont2006}.
  }
  \label{tab:surface}
  \begin{center}
    \begin{tabular}{ccccccc}
      \hline
        Date, 2011 & UT & Ocean(\%)& Vegetation(\%) & 
        TSDI(\%) & Cloud (over O+L)(\%)& Clear O/L \\
      \hline
    Mar 09 &  10:00-10:30 & 23 & 14 & 8 & 54 (26+28)& 1.05 \\
    Mar 10 &  10:00-11:30 & 19 & 14 & 9 & 56 (26+31)& 0.84 \\
    Mar 11 &  10:00-12:30 & 17 & 15 & 9 & 56 (24+32)& 0.72 \\
    Mar 12 &  11:00-14:00 & 18 & 15 & 9 & 56 (25+31)& 0.74 \\
    Mar 13 &  10:30-14:00 & 14 & 17 & 11 & 56 (23+33)& 0.51 \\
      \hline
    \end{tabular}
  \end{center}
\end{table}

\footnotetext{http://www.fourmilab.ch/earthview/vplanet.html}

\section{Data Reduction}\label{sec:reduce}
We used the HBS reduction package HBSRED (\cite{kawa1999}) for most of our data processing.
HBSRED executes the standard CCD reduction procedures and polarimetric calibration procedures, including correction for instrumental polarization. 
HBSRED is optimized for point source data;
it subtracts the sky background using the sky spectrum observed simultaneously  through  a hole next to that for the target.
This method, however, cannot be used for an extended  light source such as Earthshine.
Thus, we omitted the sky subtraction process in HBSRED and applied the following procedure.

We reproduced the portion of sky background contaminating a raw ES spectrum using the NSKY and FSKY obtained immediately after the ES observation. 
The scattered light from MS is the primary light source in the sky background around ES.
This is supported by our approximate measurement of sky brightness after moonset on March 10, which was darker than $1/30$ of NSKY.  
The intensity of the sky background increases spatially toward the MS region.
The variation can be linearly fitted as shown by the imaging observations by \citet{qiu2003} (see Figure 4 in their paper). 
Therefore, we reproduced the sky background spectrum at the ES position, $BSKY(\lambda)$, using
\begin{equation}
BSKY_{\rm{oe},\phi}(\lambda)  =  NSKY_{\rm{oe},\phi}(\lambda) 
+ \left( NSKY_{\rm{oe},\phi}(\lambda)-FSKY_{\rm{oe},\phi}(\lambda) \right) \times
\frac{r_{\rm{N}}-r_{\rm{E}}}{r_{\rm{F}}-r_{\rm{N}}}, 
\end{equation}
where $NSKY_{\rm{oe},\phi}(\lambda)$ means either an ordinary (o) or extraordinary (e)  NSKY spectrum taken with a half-wave plate angle $\phi$ of $0^{\circ}, 22.5^{\circ}, 45^{\circ},$ or $67.5^{\circ}$; 
the same applies to $FSKY_{\rm{oe},\phi}(\lambda)$ and $BSKY_{\rm{oe},\phi}(\lambda)$.
Further, $r_{\rm{E}}$, $r_{\rm{N}}$, and $r_{\rm{F}}$ represent the angular distances of the ES (11.3$'$), NSKY (20$'$), and FSKY (30$'$) positions from the Moon's center, respectively. 

\begin{figure}
  \begin{center}
    \FigureFile(80mm,80mm){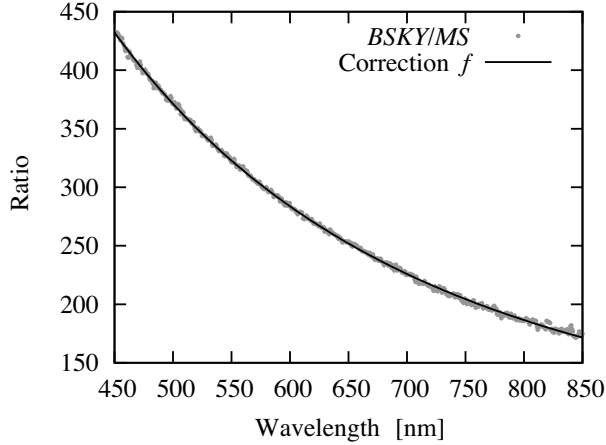}
  \end{center}
  \caption{Example of the derivation of the best-fit parameters in equation (\ref{eq:csky}). 
  The line is the best-fit correction function $f(\lambda)$ for $BSKY(\lambda)/MS(\lambda)$ data (gray dots).}
  \label{fig:msfitfig}
\end{figure}

Subtracting $BSKY_{\rm{oe},\phi}(\lambda)$ from $ES_{\rm{oe},\phi}(\lambda)$ (ES spectrum  written  using the same subscripts) would worsen the signal-to-noise (S/N) ratio.
To avoid this, $BSKY_{\rm{oe},\phi}(\lambda)$ was then replaced by $CSKY_{\rm{oe},\phi}(\lambda)$ which is a fitted spectrum with a better S/N ratio generated from the $MS_{\rm{oe},\phi}(\lambda)$ (MS spectrum) as a template with the following correction:
\begin{equation}
CSKY_{\rm{oe},\phi}(\lambda) = 
\left[ a (b \cdot \lambda - c)^{-4} + d \right] \times
MS_{\rm{oe},\phi}(\lambda)
= f(\lambda) \times MS_{\rm{oe},\phi}(\lambda),\\
\label{eq:csky}
\end{equation}
where $a$, $b$, $c$, and $d$ are obtained by the method of least squares such that $CSKY_{\rm{oe},\phi}(\lambda)$ best fits $BSKY_{\rm{oe},\phi}(\lambda)$.
The form of the correction function $f(\lambda)$ is reasonable given the wavelength dependence of telluric Rayleigh scattering of MS and stray light in the instrument.
Indeed, the $BSKY(\lambda)/MS(\lambda)$ curves are smooth and fitted excellently by $f(\lambda)$ (Fig.\ \ref{fig:msfitfig}).
Finally, $CSKY_{\rm{oe},\phi}(\lambda)$ was subtracted from $ES_{\rm{oe},\phi}(\lambda)$.

After the sky subtraction, we returned to HBSRED.
It calculates the Stokes parameters ($q=Q/I, u=U/I$) using the following equations:
\begin{eqnarray}
a_1(\lambda) &= \sqrt{ 
  \frac{ES^{'}_{\rm{e},0}(\lambda)}{ES^{'}_{\rm{o},0}(\lambda)} \times
  \frac{ES^{'}_{\rm{o},45}(\lambda)}{ES^{'}_{\rm{e},45}(\lambda)} }, \\
a_2(\lambda) &= \sqrt{
  \frac{ES^{'}_{\rm{e},22.5}(\lambda)}{ES^{'}_{\rm{o},22.5}(\lambda)} \times
  \frac{ES^{'}_{\rm{o},67.5}(\lambda)}{ES^{'}_{\rm{e},67.5}(\lambda)} },
\end{eqnarray}

\begin{eqnarray}
q(\lambda) &= \frac{1-a_1(\lambda)}{1+a_1(\lambda)}, \\
u(\lambda) &= \frac{1-a_2(\lambda)}{1+a_2(\lambda)},
\end{eqnarray}
where $ES^{'}_{\rm{oe},\phi}(\lambda)$ represents an ES spectrum after sky subtraction.
After correction of the instrumental polarization \citep{kawa1999}, the degree of polarization ($P$) and the position angle of polarization ($\alpha$) were calculated using
\begin{equation}
P(\lambda) = \sqrt{ q(\lambda)^2 + u(\lambda)^2 },
\end{equation}

\begin{equation}
\rm{tan}\ 2 \alpha(\lambda) = 
   \frac{u(\lambda)}{q(\lambda)}
   \rm{,}\ \rm{and} \  
   \rm{sgn}\left[ \rm{cos}\ 2 \alpha(\lambda)\right] = \rm{sgn}\left[q(\lambda)\right],
\end{equation}
where sgn($x$) represents the sign of $x$.

In regular (non-polarimetric, or intensity) spectroscopy of ES, an ES spectrum is divided by an MS spectrum in order to cancel the reflectance on the Moon's surface and the telluric absorption \citep{arno2008}.
However, this procedure should not be applied to spectropolarimetry of ES.
In regular  spectrometry,  the  procedure  is justified  on the  assumption  that  the  Moon reflectance that affects the ES spectra is identical to that affecting the MS spectra.  
However, this is not true in spectropolarimetry because the reflectance at each oscillation position angle depends on the phase angle. 
Note that what we derive is the polarization spectra of \textit{Earthshine} on the Moon, not \textit{Earthlight} as viewed from the space.
In the next section, we take account of the Moon reflection effect which depolarizes a polarized Earthlight spectrum.

\section{Results and Discussion}

\subsection{General Results}

\begin{figure}[p]
  \begin{center}
    \FigureFile(150mm,170mm){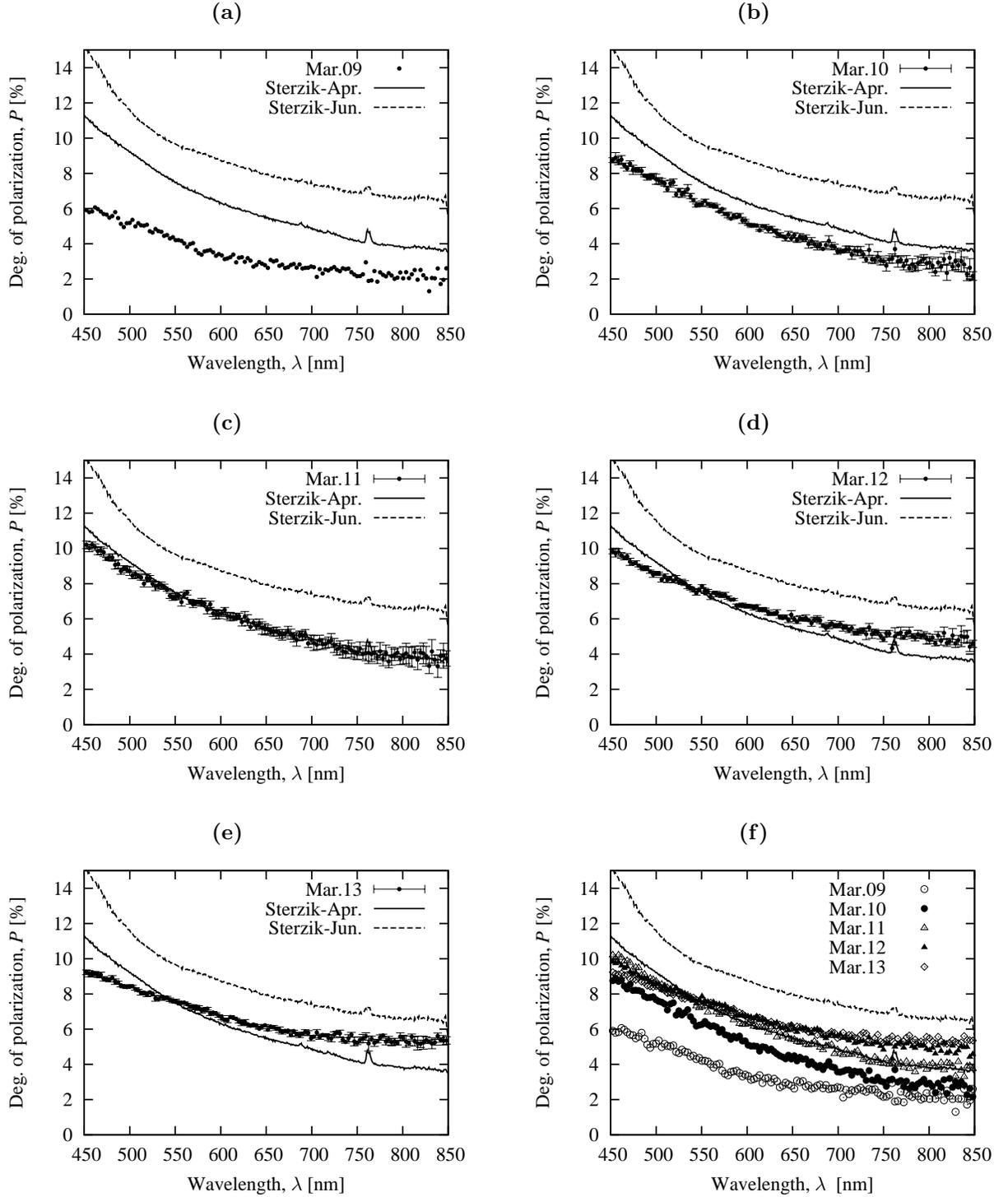}
  \end{center}
  \caption{Polarization spectra of Earthshine.
  Panels (a)-(e) are results from March 9 ($\theta$=49$^\circ$), 
  March 10 (60$^\circ$), March 11 (72$^\circ$), March 12 (84$^\circ$),
  and March 13 (96$^\circ$), respectively.
  Error bars signify standard deviation in the observed sets.
  No error bar is shown in (a) because only one effective set was obtained.
  Panel (f) is a plot of results for all the dates.
  Derived spectra are binned by 3 nm (5 pixels) to obtain a better S/N ratio.
  The results from \citet{ster2012} on 2011 April 25 ($\theta$=87$^\circ$, solid line) and 2011 June 10 (102$^\circ$, dashed line) are also plotted.}
  \label{fig:ES.P.all}
\end{figure}

The observed polarization  degree spectra are shown in Fig.\ \ref{fig:ES.P.all}.
The degree of Earthshine polarization ranges from 2\% to 10\%.
The polarization degree decreases with increasing wavelength  at  any phase.
An enhanced feature in the O$_2$  A-band at 760 nm does not appear clearly in our spectra.

\begin{figure}
  \begin{center}
    \FigureFile(80mm,80mm){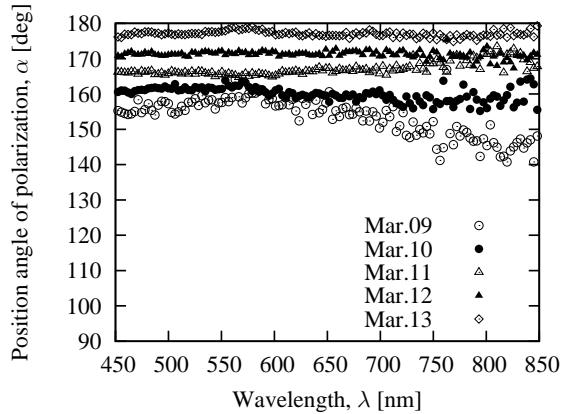}
  \end{center}
  \caption{Earthshine position angles of polarization ($\alpha$) as a function of the wavelength. $\alpha$ is taken anti-clockwise with respect to celestial north-south as viewed by an observer looking towards the source of the light beam ($0^\circ < \alpha < 180^\circ$).}
  \label{fig:ES.alf}
\end{figure}

Fig.\ \ref{fig:ES.alf} shows the observed position angles of polarization as a function of the wavelength.
The polarization position angles of a planetary reflection should be normal or parallel to the scattering plane regardless of wavelength. 
Our observational results do not show a significant dependence on the wavelength.
Table \ref{tab:alpha} summarizes the directions normal to the scattering plane ($N$) and the observed position angles averaged over the wavelength range ($\alpha$).
As expected, $\alpha$ almost coincides with $N$.
The observed position angles of polarization are reasonable.
This supports reliability of our observations and data reduction method.

\begin{table}
  \caption{Directions normal to the scattering plane ($N$) and the observed position angles of polarization ($\alpha$) averaged over 450-850 nm.}\label{tab:alpha}
  \begin{center}
    \begin{tabular}{ccc}
      \hline
      Date, 2011 & $N$ ($^\circ$) & $\alpha$ ($^\circ$) \\
      \hline
    Mar 09 & 157.4	& 154$\pm$5 \\
	Mar 10 & 162.4	& 160$\pm$2 \\
	Mar 11 & 167.8	& 167$\pm$2 \\
	Mar 12 & 173.6	& 171$\pm$1 \\
	Mar 13 & 179.3	& 177$\pm$1 \\
      \hline
    \end{tabular}
  \end{center}
\end{table}

\subsection{Comparison with Other Studies}
We compared our  results  for 2011 March 12 ($\theta=84^\circ$) and March 13 ($\theta=96^\circ$) with Earthshine spectropolarimetry on 2011 April 25 ($\theta=87^\circ$) by \citet{ster2012} (Figs.\ \ref{fig:ES.P.all}d, \ref{fig:ES.P.all}e).
The polarization degrees of our spectra have virtually the same values at 550 - 600 nm.
Our spectra decrease more slowly with  increasing wavelength  than those of
 \citet{ster2012}.
The difference between the two groups is $\sim$2 \% at the highest, which is within the range of the possible variation attributed to the changing cloud coverage \citep{wols2005}.
The spectrum on 2011 June 10 ($\theta=102^\circ$) by \citet{ster2012} exhibits higher polarization degrees than their data in April and ours on March 12-13.
This can be attributed to differences in the scene types (e.g., cloud coverage) in the Earthshine contributing area (Sterzik et al. 2012). 
They detected an enhanced polarization feature in the O$_2$ A-band at 760 nm.
It has a peak height of 0.5\%-1\%, which is comparable to the statistical error in our data.
It is generally safe to say that the observational results from the two groups are consistent. 

We compared Stam's (2008) model of an Earth-like planet  with our observation near a quadrature.
For this purpose, we needed to consider the effect of lunar reflection on the polarized Earthlight spectrum.
Lunar reflection does not add polarization because the Earth-Moon-Earth phase angle is zero \citep{coff1979}. 
Instead, it depolarizes the polarized Earthlight. 
\citet{doll1957} estimated the depolarization factor to be $\sim$3.3.
This evaluation was based on a comparison of his optical polarimetric observations of Earthshine with the roughly estimated polarization degree of Earthlight, as well as on laboratory measurements of lunar samples.
The wavelength dependence of depolarization has not been well determined.
We consider both a wavelength-independent factor of 3.3 and linearly dependent factors, i.e., $Depol = 3.3 \cdot \lambda / 550[\rm{nm}]$ (Sterzik et al. 2012).

\begin{figure}
  \begin{center}
    \FigureFile(80mm,80mm){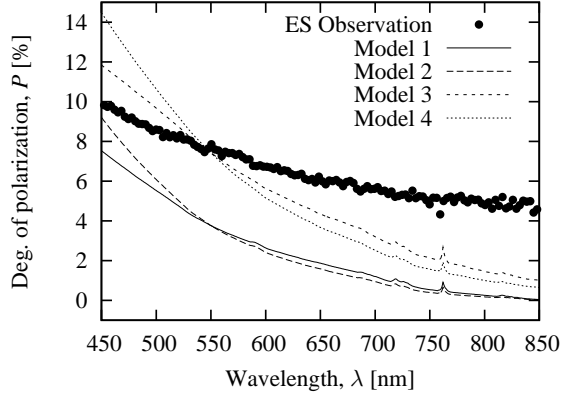}
  \end{center}
  \caption{
  Earthshine polarization spectrum on March 12 compared with the Stam's (2008) model.
  The model lines are synthesized from the templates for four different scene types and scaled by a wavelength-independent lunar depolarization factor ($Depol=3.3$) or linearly dependent factors [$Depol = 3.3 \times \lambda / 550(\rm{nm})$].
  The considered scene types are clear forest (F0), cloudy forest (F1), clear ocean (O0), and cloudy ocean (O1) [``forest'' in \citet{stam2008} means a Lambertian surface with the albedo of forest].
  The phase angle input was set to be 84$^{\circ}$, which fits the observation.
 Model 1 employs a parameter setting close to the exact circumstances (Table \ref{tab:surface}), i.e., $\textrm{F0}:\textrm{F1}:\textrm{O0}:\textrm{O1}= 24:31:18:25$ and is scaled by the wavelength-independent depolarization factor.
 Model 2 is the same as Model 1 except that it is scaled by the linearly wavelength-dependent factors.
 Models 3 and 4 are the best-fit models within the entire wavelength range, obtained by varying the ocean/land ratio (from 0.1 to 10) and cloud coverage (from 0 to 1).
 Model 3 is scaled by the wavelength-independent factor.
 Model 4 is scaled by the linearly dependent factors.
 The resulting parameter setting for both models is $\textrm{F0}:\textrm{F1}:\textrm{O0}:\textrm{O1}=6:3:64:27$.}
  \label{fig:compmodel}
\end{figure}

Fig.\ \ref{fig:compmodel} shows the Earthshine polarization spectrum observed  on March 12  ($\theta=84^\circ$), compared with the Stam's (2008) model spectra for four sets of the parameters.
Although  both the observed and modeled spectra exhibit decreasing polarization degree with increasing wavelength, the  model spectra do not  match  those  based on the observation.
\citet{ster2012} made a similar analysis and also found considerable differences between their observations and  the  model.
Because the Earthshine observations by the two groups are consistent, further model sophistication and/or re-examination of the lunar depolarization factor is expected to resolve this discrepancy.

\subsection{Phase Variations}\label{sec:phase}
\begin{figure}
  \begin{center}
    \FigureFile(80mm,80mm){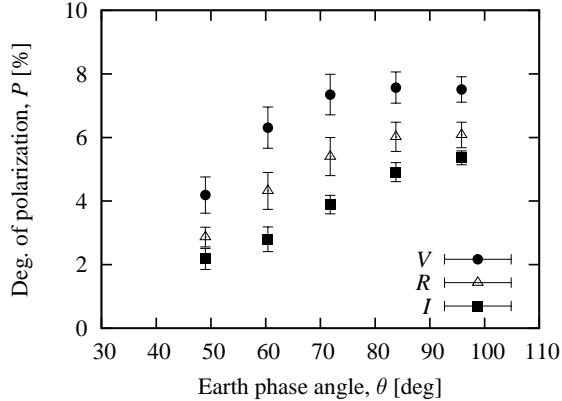}
  \end{center}
  \caption{Earthshine degree of polarization as a function of the Earth phase angle, averaged for the $V$ (507-599 nm), $R$ (589-727 nm), and $I$ (731-881 nm) bands.}
  \label{fig:phaseP}
\end{figure}

Fig.\ \ref{fig:ES.P.all}f shows that the overall degree of polarization increases as the Earth phase approaches quadrature.
Interestingly, with a more detailed look, the phase dependence appears to differ with the wavelength.
To express this, we average the degree of polarization in the $V$, $R$, and $I$ bands and plot them against the Earth phase angle (Fig.\ \ref{fig:phaseP}).
The result suggests that $\theta_{\rm{max}}$ (the phase angle at which the maximum polarization degree occurs) is larger for longer wavelengths: 
$\theta_{\rm{max}}$ for the $V$ band is $\sim 90^\circ$, whereas $\theta_{\rm{max}}$ for the $I$ band seems to be larger than $96^\circ$.

$\theta_{\rm{max}}$ varies depending on the type of scattering/reflection process.
As described in section \ \ref{sec:intro}, the polarization of Earthlight (thus of Earthshine) is caused by scattering in the atmosphere and surface reflection.
Polarization from Rayleigh scattering by atmospheric molecules has 
$\theta_{\rm{max}}$ at 90$^\circ$ \citep{hapk1993}. 
On the other hand, $\theta_{\rm{max}}$ for surface reflection can be larger than 90$^\circ$.
The total light reflected from a particulate surface can be understood as the sum of three components: surface singly scattered rays, volume-scattered rays, and multiply scattered rays \citep{woes1987}.
Because the volume- and multiply scattered rays are generally only weakly polarized, the primary source of polarization in the total surface reflection is the surface singly scattered light, which can be expressed approximately by the Fresnel equations for specular reflection \citep{woes1987}.
Typical refractive indices for soils, vegetation, and liquid water are $\sim$1.55 \citep{woes1987,eshe2004}, $\sim$1.38 \citep{woes1987,jone2010}, and $\sim$1.33 \citep{daim2007}, for which the Fresnel equations yield $\theta_{\rm{max}}$  values of $\sim$114$^{\circ}$, $\sim$108$^{\circ}$, and $\sim$106$^{\circ}$, respectively. 
Actual measurements proved that $\theta_{\rm{max}}$ can be larger than 90$^\circ$ for surface reflection:
for soils, polarimetry of a terrestrial soil (together with desert sand), the Moon and Mercury (i.e., airless solid bodies) revealed that their $\theta_{\rm{max}}$ values are roughly 100-110$^\circ$\citep{chen1968}, 95-110$^\circ$ \citep{coyn1970, doll1971, hapk1993}, and 93-103$^\circ$ \citep{gehr1987}, respectively;
for vegetation, $\theta_{\rm{max}}$ values of 90-120$^\circ$ or beyond were measured by \citet{woes1987} and \citet{suom2009};
for liquid water, a $\theta_{\rm{max}}$ of $\sim$106$^\circ$ was reported by \citet{chen1968}.

These estimates and measurements suggest that the Earthshine polarization at  shorter wavelengths is dominated by Rayleigh scattering by atmospheric molecules, whereas that at longer wavelengths has an increasingly effective contribution from reflection at the land and ocean surfaces of the Earth.
Indeed, considering the wavelength dependences of the scattering/reflectance processes, the fraction of the atmospheric Rayleigh scattering in the total Earthlight should be much larger at shorter wavelengths than it is at  longer wavelengths:
the Rayleigh scattering radiance has a strong negative correlation (proportional to $1/\lambda^4$), whereas the reflectances of soils and vegetation have generally positive correlations\footnote{Regarding vegetation,  a more detailed  description  is that its  reflectance  is comparatively constant at 300-700 nm and jumps steeply around  700 nm, which is called the vegetation red edge.}; that of liquid water has a linearly (i.e., slower than Rayleigh scattering) negative correlation, according to the ASTER Spectral Library\footnote{http://speclib.jpl.nasa.gov/}.

As additional processes for  producing  the observed wavelength dependence of $\theta_{\rm{max}}$, we discuss the changes in scene (ocean/land ratio, cloud coverage) in the Earthshine-contributing area on the Earth during our observing period.
Since the cloud coverage was virtually constant throughout the observation period (Table \ref{tab:surface}), it should not be effective. 
The ocean/land ratio decreased toward the end of the observations (Table \ref{tab:surface}).
As described, the estimated or measured $\theta_{\rm{max}}$ for both liquid water and land (soils and vegetation) were typically greater than 90$^\circ$.
Hence, the $\theta_{\rm{max}}$ of $\sim$90$^\circ$ cannot be explained without atmospheric Rayleigh scattering.
A decrease in the ocean fraction will increase the ratio of the surface (ocean and land) reflection to the atmospheric Rayleigh scattering in the total Earthlight because the land is brighter than the ocean, and thus may produce a larger $\theta_{\rm{max}}$.
This effect should be more apparent at longer wavelengths, given the wavelength dependences of atmospheric Rayleigh scattering (strongly negative) and surface reflection (positive or weekly negative).
Hence, a decrease in the ocean/land ratio is at least qualitatively capable of producing the observed wavelength dependence of $\theta_{\rm{max}}$.
Note that this process is consistent with the interpretation described in the previous paragraph because the determining factor in both  discussions is the same:  the wavelength  dependences of atmospheric  Rayleigh scattering  and  surface reflection.

To summarize the discussion of the observed wavelength dependence of phase variations in Earthshine polarized spectra, the results should be interpreted as an outcome of the wavelength-dependent ratio of atmospheric Rayleigh scattering to surface reflection.
Earthshine polarization at shorter wavelengths is dominated by atmospheric Rayleigh scattering, whereas that at longer wavelengths has an increasingly effective contribution from surface reflection.
It is qualitatively possible that a decrease in the ocean fraction in the Earthshine-contributing area was an additional process contributing to the observed wavelength dependence of the phase variation.

\subsection{Implications}
The polarization  of Earthlight, which is estimated from the Earthshine polarized spectrum, has implications concerning the search for Earth-like  extrasolar planets. 
Assuming a depolarization factor of $\sim$3.3 at the lunar surface \citep{doll1957}, the polarization  degree of Earthlight should be as much as $\sim$30\% at 450 nm and $\sim$20\% at 850 nm when the Earth  phase is near quadrature.
On the other hand, airless solid planetary bodies have a lower degree of the maximum polarization; 
maximum polarizations of 6\%-22\% at 435 nm and 5\%-9\% at 840 nm were measured for the Moon \citep{doll1971}, 
and values of 9\%-10\% at 450 nm and 6\%-7\% at 850 nm were measured for Mercury \citep{gehr1987}.
Our observations of Moonshine are also within the ranges measured by \citet{doll1971}. 
These results suggest that Earth-like planets have significantly higher polarization than do airless solid bodies.

The wavelength dependence of $\theta_{\rm{max}}$, the phase angle of the maximum polarization (subsection.\ \ref{sec:phase}), may also be used as a signature of Earth-like planets.
For Earthshine, and thus for Earthlight, $\theta_{\rm{max}}$ increases with increasing wavelength:
$\theta_{\rm{max}}$ in the $V$ band is a $\sim 90^\circ$ phase angle, whereas that for longer wavelengths appears at larger phase angles. 
In contrast, $\theta_{\rm{max}}$ for airless solid bodies, namely, the Moon and Mercury, decreases with increasing wavelength (\cite{coyn1970}; Fig.\,19 in \cite{doll1971}; Table VI in \cite{gehr1987}). 

Finally,  we compare our results with those for other  types  of terrestrial  planets  than  airless  solid bodies.
Venus,  a  terrestrial planet  covered  with  thick  atmospheric clouds,   exhibits  a completely different polarimetric behavior, which is explained by the Mie theory for cloud particles with a refractive index of $\sim$1.44 and a mean radius of $\sim$ 1.1 $\micron$ \citep{hans1974}:
``negative'' polarization (i.e., polarization  in the  direction parallel to  the  scattering  plane) is observed at most visible wavelengths and 30-140$^\circ$ phase angles;
$\theta_{\rm{max}}$ is located around 15-20$^\circ$ phase angles in the 350-600 nm region \citep{doll1970}.
Therefore, Venus-like planets should be distinguishable from Earth-like planets.

For Mars, a terrestrial planet with a thin atmosphere, we cannot access any near-full-phase polarimetric measurements in multiple wavelengths.
\citet{doll1983} presented polarization measurements by the spacecraft Mars-5 at phase angles of 60$^\circ$ and 90$^\circ$ at a wavelength of 592 nm, together with ground-based observations for phase angles smaller than 43$^\circ$.
The results generally resembled the polarization of the Moon and Mercury:
the maximum polarization was extrapolated to be $\sim$7\% at a phase angle of $\sim$100$^\circ$.
The maximum polarization of Mars is significantly smaller than that of the Earth.
Its wavelength dependence of $\theta_{\rm{max}}$ is inferred to be similar to that of the Moon, which has an opposite dependence to that of the Earth.

Note that uncertainty in the lunar depolarization factor may affect the estimated polarization degree of Earthlight. 
However, it does not affect $\theta_{\rm{max}}$; 
the reason is that the depolarization  factor at a certain wavelength should be conserved regardless of the Earth  phase angles because the Earth-Moon-Earth angle is constant (always zero).

\section{Conclusions}
We conducted spectropolarimetry of Earthshine for five Earth phase angles ranging from 49$^{\circ}$ to 96$^{\circ}$.
The  observed spectra exhibit a decreasing Earthshine polarization  degree with increasing wavelength at any phase. 
The overall degree of polarization increases as the Earth approaches a quadrature phase.
The observed phase dependence varies with the wavelength; 
the maximum polarization for the $V$ band occurs at a phase angle of $\sim 90^\circ$, whereas that for longer wavelengths is reached at larger phase angles.
% Add on 2012-10-28
Further observations at phase angles larger than 100$^\circ$ are desirable to establish the exact $\theta_{\rm{max}}$ at  longer wavelengths.
% was "This is" 
The observed wavelength dependence of $\theta_{\rm{max}}$ is interpreted as indicating that the Earthshine polarization at shorter wavelengths is dominated by atmospheric Rayleigh scattering, whereas that at longer wavelengths has an increasingly effective contribution from the Earth surface reflection.
The wavelength dependence of $\theta_{\rm{max}}$ for the Earth appears to be unique among the terrestrial planetary bodies in the Solar System.
Therefore it might be a signature expressing a distinctive characteristic  of the Earth: this planet has a scattering but transparent atmosphere above its surface.

\bigskip

We thank the staff members of Okayama Astrophysical Observatory, National Astronomical Observatory of Japan.
We are deeply grateful to Dr.~Enric Pall{\'e}, the referee, for giving helpful comments as well as kindly calculating the Earth scenery data in Table \ref{tab:surface}.
We also acknowledge Dr.~Michael F. Sterzik for providing us the observational data for Fig.\ \ref{fig:ES.P.all}. 
Part of this study was supported by the Center for Planetary Science 
running under the auspices of the MEXT GCOE Program entitled 
``Foundation of International Center for Planetary Science'', a joint project between Kobe University and Hokkaido University.
This work was also supported by JSPS KAKENHI Grant Number 24540231.

%\newpage

\end{document}